\documentclass[fleqn,10pt]{wlscirep}
\usepackage[T1]{fontenc}
\usepackage{color}
\usepackage[normalem]{ulem}

\title{Simulating polaron biophysics with Rydberg atoms}
 
\author[1,2,*]{Marcin P\l{}odzie\'n}
\author[2]{Tomasz Sowi\'nski}
\author[1]{ Servaas Kokkelmans}
\affil[1]{Department of Applied Physics, Eindhoven University of Technology, PO Box 513, 5600 MB Eindhoven, The Netherlands}
\affil[2]{Institute of Physics, Polish Academy of Sciences, Aleja Lotnik\'{o}w 32/46, PL-02668 Warsaw, Poland}
\affil[*]{marcin.plodzien@ifpan.edu.pl}

\begin{abstract}
Transport of excitations along proteins can be formulated in a quantum physics context, based on the periodicity and vibrational modes of the structures.  Numerically exact solutions of the corresponding equations are very challenging to obtain on classical computers. Approximate solutions based on the Davydov ansatz have demonstrated the possibility of stabilized solitonic excitations along the protein, however, experimentally these solutions have never been directly observed. Here we propose an alternative study of biophysical transport phenomena based on a quantum simulator composed of a chain of ultracold dressed Rydberg atoms, which allows for a direct observation of the Davydov phenomena. We show that there is an experimentally accessible range of parameters where the system directly mimics the Davydov equations and their solutions. Moreover, we show that such a quantum simulator has access to the regime in between the small and large polaron regimes, which cannot be described perturbatively. 
\end{abstract}
\begin{document}

\flushbottom
\maketitle
\thispagestyle{empty} 

\section*{Introduction}
The understanding of molecular structures provides life sciences with tools to explain complex cell-biology phenomena. Biological complexity of mesoscopic objects along with the quantum behavior of their basic elements leads to interesting unsolved questions awaiting comprehensive answers \cite{Whaley2010}.  The interdisciplinary field of ``quantum biology''  is the natural area for combining quantum physical methods and tools to investigate, model and simulate biological systems on a mesoscopic level \cite{Arndt2009,Plenio2013,Li2016}.
 
Many biological processes are powered by the energy released from the hydrolysis of adenosine triphosphate (ATP). On a physical level, it can be viewed as a  vibrational bound state of the ATP molecule to a  protein with energy equal to $0.49$ eV. In the 1970's Davydov proposed a mechanism for the localization and transport of the associated vibrational energy in the  $\alpha$-helix region of a protein by means of a so-called  Davydov soliton \cite{Davydov1969,Davydov1973a,Davydov1973b,Davydov1976}. 
%The vibrational degrees of freedom are coupled to excitons %forming  an exciton-vibration localized state, called the %Davydov's soliton. 
Although this model has been used for a theoretical description of experimentally observed unconventional absorption bands in proteins \cite{Scott1983,Scott1984}, direct experimental evidence for existence of this soliton is still missing. 

The Davydov soliton is a subclass of  richer  {\it polaron}  phenomena, \textit{i.e}.,  excitations  mediated by phonons originally introduced by Landau \cite{Landau}. Polarons have been broadly studied theoretically as well as experimentally in a condensed matter context \cite{Alexandrov} and recently in different areas of  ultracold physics: ultracold ions \cite{Cirac2004,Zoller2008,Monroe2009,Hague2012,Solano2012,Cirac2012,Lamata2014}, polar molecules \cite{Herrera2010a,Herrera2010b,Herrera2011,Lesanovsky2012,Herrera2013}, ultracold Rydberg gases \cite{MacCormick2012,Hague2014,Wuster2015,Graetzle2015,Pohl2015,Molmer2017}, and  strongly-interacting ultracold Bose and Fermi gases  \cite{Zwerlein2009,Kohl2012,Deborah2016,Jorgensen2016,Levinsen2016,Nakano2017,Demler2017,Roati2017}.  
In this paper we show that the Davydov soliton can be created and observed in a suitably prepared system of ultracold atoms, confined in an optical lattice and off-resonantly coupled to a  Rydberg state \cite{Henkel2010,Johnson2010,Honer2010,Pupillo2010}. Such a system can be regarded as a dedicated quantum simulator within the broader class of the Holstein-Su-Schrieffer-Heeger (HSSH) Hamiltonian \cite{Holstein,SSH_model}.
 First, we investigate dynamical properties of the system within the semi-classical Davydov approach assuming an infinite number of phonons. Second, we use an exact evolution approach to study the dynamics of the system in the non-classical few-phonons regime. In a particular, experimentally accessible parameter regime, both approaches confirm the existence of soliton solutions. This also indicates that with this Rydberg quantum simulator it is possible to study the regime in between the so-called small and large polaron regimes of the HSSH model, which cannot be described by perturbative theoretical methods.  
 
\section{The simulator}
 We mimic the behavior of the above-mentioned bio-molecules with a system of ultracold atoms confined in a very deep one-dimensional optical lattice potential $V(x)=V_0\sin^2(2\pi x/R_0)$ where each lattice site is occupied exactly by one atom. We assume that the spatial dynamics of atoms is not completely frozen, \textit{i.e.}, atoms may oscillate in vicinities of local minima with frequency $\omega_0 = \sqrt{2V_0\pi^2/mR_0^2}$. This motion is however quantized and therefore it is driven by a simple harmonic oscillator-like Hamiltonian:
\begin{equation}
\hat{\cal H}_{\text{vib}} = \sum_i\left(\frac{\hat p_i^2}{2m}+\frac{m\omega_0^2}{2}\hat{u}_i^2\right) = \sum_i \hbar\omega_0 \hat{b}_i^\dagger\hat{b}_i,
\end{equation} 
where $\hat{u}_i = l_0(\hat{b}^\dagger_i + \hat{b}_i)/\sqrt{2}$ and $\hat{p}_i = i\hbar(\hat{b}^\dagger_i - \hat{b}_i)/(l_0\sqrt{2})$ are the position and momenta operators related  to $i$-th atom, while operator $\hat{b}_i$ annihilates vibrational excitation of the $i$-th atom. Local motion defines a natural scale of length, $l_0 = \sqrt{\hbar/m\omega_0}$.

Besides spatial motion, each atom may exhibit changes of its internal state due to the long-range interactions between neighboring atoms via Rydberg dressing mechanism \cite{Pohl2014,Buchler2010,Rost2014,Gross2016,Ates2008}. Following the work by W\"{u}ster {\it et al.}\cite{Wuster2011} coupling of the internal state of atoms with their spatial motion can be realized by the off-resonant coupling of two different but degenerated internal Zeeman levels in the different hyperfine states of the ground-state manifold of the atoms. The large hyperfine splitting will permit selective addressing of the levels $|g\rangle$ and $|g'\rangle$ to two precisely selected, highly excited Rydberg states $|nS\rangle$ or $|nP\rangle$ (via two- and single-photon transition, respectively) with principal quantum number $n$ and angular momentum equal to $0$ or $\hbar$, respectively. A perturbation analysis shows that this coupling results in a quite small admixture of a Rydberg state to the atomic ground states and, as a consequence, atom can be found in one of the two dressed states\cite{Wuster2011}:
\begin{align}
|\mathtt{0}\rangle \approx  |g\rangle  + \alpha_s  |nS\rangle, \qquad 
|\mathtt{1}\rangle \approx  |g'\rangle + \alpha_p  |nP\rangle,
\end{align}
where amplitudes $\alpha_{l}  = \varTheta_{l}/2\varDelta_{l}$ ($l\in\{s,p\}$) are determined by a total Rabi frequency of a driving field $\varTheta_{l}$ and a total laser detuning $\varDelta_{l}$. 
In this basis of dressed states the dipole-dipole interaction between neighboring atoms is $C_3^{sp}/R^3$ ($R$ is the spatial distance between the atoms), besides additional contribution to the energy gap between local states $|\mathtt{0}_i\rangle$ and $|\mathtt{1}_i\rangle$, may induce transitions  (excitation hoppings)  between internal states of neighboring atoms  $|\mathtt{0}_i\rangle|\mathtt{1}_{i+1}\rangle\leftrightarrow |\mathtt{1}_i\rangle|\mathtt{0}_{i+1}\rangle$. In consequence, the excitation $|\mathtt{1}\rangle$ can be effectively transported across the lattice. This effect is driven by the following Hamiltonian of internal motion of all atoms:
\begin{equation}\label{Hamiltonian_1}
\hat{\cal H}_\mathrm{exc}  = \sum_i W_i\hat{a}^\dagger_i\hat{a}_i + \sum_iJ_{i+1,i}(\hat{a}^\dagger_{i+1}\hat{a}_i+\hat{a}^\dagger_{i}\hat{a}_{i+1}),
\end{equation}
where an annihilation operator of an excitation $\hat a_i$ can be viewed as a local transition operator $|\mathtt{1}_i\rangle\langle\mathtt{0}_i|$ between dressed Rydberg states. The  spatial dependent  parameters $W_i$ and $J_{i,i+1}$ are related to the dipole-dipole forces induced by Rydberg dressing and are given by\cite{Wuster2011}:
\begin{align}\label{JiW}
W_i  &=\frac{\alpha^4\hbar\varDelta }{2}\left(\frac{1}{1-\kappa(R_0+u_{i+1}-u_i)^2} + \frac{1}{1-\kappa(R_0+u_{i}-u_{i-1})^2}\right), \nonumber \\
J_{i+1,i} &=\frac{\alpha^4 C_3^{sp}}{|R_0+u_{i+1}-u_i|^3} \frac{1}{1-\kappa(R_0+u_{i+1}-u_i)^2}, 
\end{align}
where  $\kappa(R) = (C_3^{sp}/\hbar\varDelta)/ R^3$ and $\varDelta = \varDelta_s + \varDelta_p$.
In a static situation, when all atom positions are frozen, the energies $W_i$ and $J_{i+1,i}$ are site-independent with values controlled by dipole-dipole interactions between neighboring atoms at fixed  lattice spacing    $R_0$.  However, due to the vibrational motion of atoms, these parameters are position dependent and they couple internal states of atoms with their motional degrees of freedom. In the lowest order of approximation they can be written as $W_i = W_0 + g_W(u_{i+1}+ u_{i-1})$, $J_{i+1,i}  = -J_0 + g_J (u_{i+1}- u_i)$,  where $g_W$ and $g_J$ are  the appropriate Taylor expansion coefficients   of \eqref{JiW} around $R_0$.
Moreover, since the vibrational motion is quantized, the parameters have an operator character when acting in the subspace of spatial motion of atoms. 
By inserting expanded $W_i$ and $J_{i+1,i}$ to the Hamiltonian \eqref{Hamiltonian_1} one obtains HSSH Hamiltonian
% \begin{equation}\label{Hamiltonian_final}
%  \cal{\hat{H}} = \cal{\hat{H}}_\mathrm{exc} + \cal{\hat{H}}_{\mathrm{vib}},
% \end{equation}
% \textcolor{blue}{
% \begin{equation}\label{Hamiltonian_final}
%  \cal{\hat{H}} = \cal{\hat{H}}_\mathrm{0} + \cal{\hat{H}}_{\mathrm{W}} + \cal{\hat{H}}_{\mathrm{J}},
% \end{equation}

\begin{equation}\label{Hamiltonian_final}
\begin{split}
 \cal{\hat{H}} = & \sum_i W_0 \hat{a}^\dagger_i\hat{a}_i -J_0(\hat{a}^\dagger_{i+1}\hat{a}_i + h.c.) +  \sum_i  g_W(u_{i+1} - u_{i-1})\hat{a}^\dagger_i\hat{a}_i + \sum_i  g_J(u_{i+1} - u_{i})(\hat{a}^\dagger_{i+1}\hat{a}_i + h.c.)
\end{split}
\end{equation}
where the first term describes the excitation dynamics on the lattice, the second term describes the excitation-vibration coupling via the on-site energy, and the last term represents the off-site coupling via the excitation-vibration coupling through the hopping energy. Our implementation can be regarded as a dedicated quantum simulator to study excitation stabilization by vibrations related to the $\alpha$-helix protein. 
For the Rydberg parameters that we consider, the second order terms in the Taylor expansion are significantly smaller than the first-order corrections, which justifies a linear approximation.

For the moment we comment on a special case of Hamiltonian (\ref{Hamiltonian_final}) with zero off-site coupling $g_J = 0$, \textit{i.e.} Holstein model. In this model the excitation is dressed by phonons forming a polaron quasiparticle. Two limiting cases have exact solutions: $(i)$ For zero excitation-vibration coupling  $g_W = 0$ eigenstates are well described by product states in Fourier space \textit{i.e.} $\hat{a}^{\dagger}_k|0\rangle, \hat{a}^\dagger_{k-q}\hat{b}^\dagger_q|0\rangle, \hat{a}^\dagger_{k-q-q'}\hat{b}^\dagger_q \hat{b}^\dagger_{q'}|0\rangle$, etc. These states construct a good basis for perturbation expansions in the $g_\text{W}/J_0$ parameter for the weak coupling limit; $(ii)$ a second limiting case, called the small polaron, corresponds to a zero hopping energy term $J_0 = 0$ in which the Hamiltonian can be diagonalized in the dressed-polaron picture, \textit{i.e.}
\begin{equation}
\begin{split}
  \cal{\hat{H}}   & =  \sum_{i}\hbar\omega_0(\hat{b}_{i}^\dagger \hat{b}_{i}) + g_w  \sum_{i}  \hat{a}_i^\dagger a_i (\hat{b}_{i}^\dagger + \hat{b}_{i}) = \hbar\omega_0\hat{b}_0^\dagger\hat{b}_0 + g_W(\hat{b}_0^\dagger+\hat{b}_0) + \hbar\omega_0\sum_{i\neq 0}(\hat{b}_i^\dagger\hat{b}_i+1/2)  \\
      & = \hbar\omega_0\hat{B}_0^\dagger\hat{B}_0 - \frac{g_W^2}{\hbar\omega_0} + \hbar\omega_0\sum_{i\neq 0}(\hat{b}_i^\dagger\hat{b}_i),
  \end{split}
\end{equation}
where $\hat{B}_0 = \hat{b}_0 +  g_W/\hbar\omega_0$. The dressed-polaron eigenstates $\hat{a}^\dagger_0 |0\rangle_{ex}\hat{B}^\dagger_0 \prod_{i\neq 0}\hat{b}^{\dagger^{m_i}}|\tilde{0}\rangle_{ph}$ with eigenenergies 
$E_n=-g_W^2/\hbar\omega_0 + n\hbar\omega_0$ form a good basis for a perturbative description in the small $g_W/\hbar\omega_0$ parameter .
A perturbative calculation shows that the excitation-vibration coupling is given by a dressed hopping amplitude $J_0e^{-(g_W/\hbar\omega_0)^2}$\cite{Alexandrov},  where for $\omega_0$ large enough, the hopping amplitude vanishes. For the HSSH model in the regime of parameters giving rise to a Davydov soliton, \textit{i.e.} where all parameters are of the same order, a good small parameter for a perturbative description is lacking, and therefore a variational approach is preferred.
 
 The Rydberg dressing is responsible for the coupling between vibrational degrees of freedom of neighboring atoms. The resulting dressed soft-core interaction is proportional to $(R^6+R_b^6)^{-1}$, with $R_b$ the Rydberg blockade radius \cite{Honer2010}, gives also rise to an additional energy shift. However, this shift is negligible compared to $\hbar\omega_0$, and therefore we omit it. In the following, all energies are expressed in units of $J_0$, and time is measured in units of $\hbar/J_0$, {\it i.e.} we set $J_0=\hbar=m=1$. 

\section{Dynamical properties of the system}
An important question related to the dynamics of the HSSH Hamiltonian is whether the lattice vibrations are able to stabilize the excitation that is initially localized on a specific site $K$ 
$
|\boldsymbol{\Psi}_0\rangle = \hat{a}_K^\dagger|\mathtt{vac}\rangle,
$
or  slightly delocalized  on two  neighboring sites
$
|\tilde{\boldsymbol{\Psi}}_0\rangle = \frac{1}{\sqrt{2}}(\hat{a}_K^\dagger + \hat{a}_{K+1}^\dagger)|\mathtt{vac}\rangle,
$
where $|\mathtt{vac}\rangle$ is the vacuum state of the system fulfilling the condition $\hat{a}_i|\mathtt{vac}\rangle=\hat{b}_i|\mathtt{vac}\rangle=0$ for any $i$. 
For certain parameters, a system prepared in these initial states evolves in such a way that the excitation does not spread across the protein. This is attributed to a specific ratio of the  interactions of excitation and vibrational degrees of freedom, giving rise to a soliton.  This  spreading or non-spreading behavior can be extracted from information encoded in the time-dependent density profile $\rho_i(t) = \langle\boldsymbol{\Psi}(t)|\hat{a}^\dagger_i\hat{a}_i|\boldsymbol{\Psi}(t)\rangle,
$
where the state of the system at given time $t$  can be formally written as
%\begin{equation} \label{Evolution}
$|\boldsymbol{\Psi}(t)\rangle = \mathrm{exp}\left(-i\hat{\cal H}t\right)|\boldsymbol{\Psi}_\mathtt{ini}\rangle$,
%\end{equation}
where $|\boldsymbol{\Psi}_\mathtt{ini}\rangle$ is one of  the  considered initial states.
% $
% |\boldsymbol{\Psi}(t)\rangle = \mathrm{exp}\left(-i\hat{\cal H}t\right)|\boldsymbol{\Psi}_0\rangle.
% $
Temporal spreading of the excitation is  then given by an effective width of the spatial density profile
$
\sigma(t) = (N\left[\sum_i \rho_i^2(t)\right])^{-1}.
$
This quantity takes the value $1/N$ for an excitation localized at exactly one lattice site and $1$ when fully delocalized. In principle, by analyzing the time-dependence of $\sigma(t)$ one can easily determine whether  the excitation remains localized or whether it spreads across the system. Numerically exact solutions of the evolution problem 
% \eqref{Evolution} 
are very challenging   due to the strong non-linear quantum-mechanical coupling between excitation and vibrational degrees of freedom. Therefore generally the evolution of the system cannot be found exactly  and some  approximation  methods have to be adopted. 

\subsection*{The Davydov approach}

We discuss here the two-step Davydov approach \cite{Davydov1969}, which results in a semiclassical description of the system. In the first step one assumes that the state of the system $|\boldsymbol{\Psi}(t)\rangle$ can be well approximated by the product of two independent states $|\psi(t)\rangle$ and $|\phi(t)\rangle$ for excitation and vibrational degrees of freedom, respectively, $
|\boldsymbol{\Psi}(t)\rangle = |\psi(t)\rangle|\phi(t)\rangle$. Since the system is initially prepared in the state with precisely one excitation and the number of excitations is conserved, the state $|\psi(t)\rangle$  can be decomposed  in the single-particle subspace, $|\psi(t)\rangle=\sum_i\psi_i(t)\hat{a}_i^\dagger|\mathtt{vac}\rangle$, where time-dependent functions $\psi_i(t)$ play   the  role of probability amplitudes  for  finding an excitation at site $i$. Consequently $\rho_i(t)=|\psi_i(t)|^2$. The second step relays on a semi-classical treatment of the vibrational degrees of the system. In analogy to other quantum field theories, we assume that the state $|\phi(t)\rangle$ has   classical features, {\it i.e.}, it can be well approximated by the product of independent coherent states:
$
|\phi(t)\rangle =\mathrm{exp}\left[-i\sum_i( u_i(t)\hat{p}_i-p_i(t)\hat{u}_i)\right]|\mathtt{vac}\rangle,
$
where amplitudes $u_i(t)$ and $p_i(t)$ are expectation values of appropriate operators in the state $|\phi(t)\rangle$. Within these approximations,  we calculate the expectation value of the many-body Hamiltonian on the system state and approximate the resulting equation of motion by the classical Hamilton equations \cite{Zhang1988}, we obtain
%
% it can be shown straightforwardly that the evolution 
%equation \eqref{Evolution} 
%is equivalent to the 
set of coupled differential equations of the form:
\begin{equation} \label{DavidovEq}
\begin{aligned}
i\frac{\mathrm{d}\psi_i(t)}{\mathrm{d}t} & = -(\psi_{i+1} + \psi_{i-1}) + g_W(u_{i+1} - u_{i-1})\psi_i  + g_J[\psi_{i+1}(u_{i+1} - u_i) + \psi_{i-1}(u_i - u_{i-1})],  \\
\frac{\mathrm{d}u^2_i(t)}{\mathrm{d}t^2} &= -\omega_0^2 u_i(t) +  g_W \omega_0(|\psi_{i+1}|^2 - |\psi_{i-1}|^2)  +  g_J \omega_0[\psi_i^*(\psi_{i+1} - \psi_{i-1}) + \psi_i(\psi_{i+1}^* - \psi_{i-1}^*)]. 
\end{aligned}
\end{equation}
These are Davydov type equations \cite{Davydov1969,Davydov1973a,Davydov1973b,Davydov1976}, which 
describe the dynamics of an excitation $\psi_i$ coupled to a gradient of a classical  phonon  field $u_i$ forming an effective self-trapping potential. An alternative derivation provided by Kerr \cite{Kerr1987} is based on the Heisenberg equations of phonon position and momentum operators. A complementary approach to the above Davydov equations, which are based on the Lagrangian variational principle, is one based on the Dirac-Frenkel-McLachan (DFM) variational principle.
This approach is commonly used in quantum molecular dynamics \cite{McLachlan1964,Lubich2004,Raab2000,Zhao2004,Perroni2004,Stojanovic2008,Zhou2012},  in which equations of motion for the variational parameters are obtained from the minimization of $\langle\delta\phi|\hat{\cal H}-i\partial_t|\phi\rangle$, where $\delta\phi$ denotes possible 
variations of $\phi$ with respect to the variational parameters. 
 
\subsection*{Phase diagram}
\begin{figure}
  \includegraphics[scale=0.3]{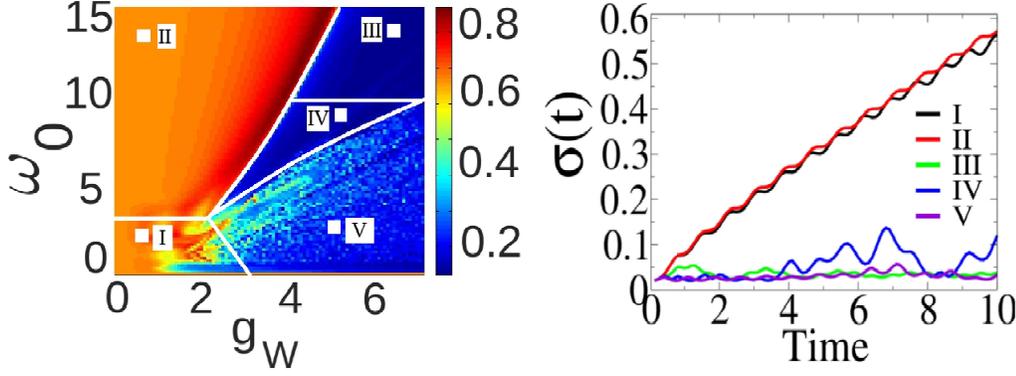} 
 \caption{(Color online) Left panel: Maximal value of an effective width of the spatial density profile $\mathrm{max}\left[\sigma(t)\right]$ as a function of $\omega_0$ and $g_W$ for vanishing coupling $g_J = 0$. A sharp crossover between non-spreading excitations (blue) and spreading excitation (dark red) is clearly visible. Different regions of the phase-diagram (bordered with white lines) correspond to a distinct nature of the exciton-vibration dynamics. 
Right panel: Evolution of the excitation width $\sigma(t)$ for different points on the phase diagram (marked as white squares on the left panel).}
\label{fig1}
\end{figure}

\begin{figure} 
\includegraphics[scale=0.45]{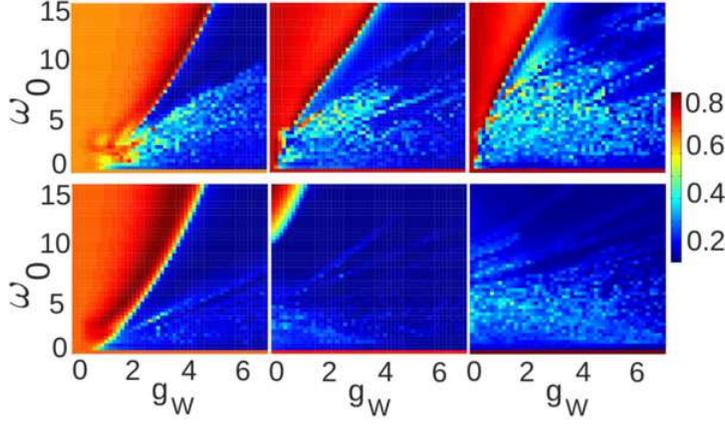} 
\caption{(Color scale) Maximal value of the wave packet width $\mathrm{max}\left[\sigma(t)\right]$ for different initial states $|\boldsymbol{\Psi}_0\rangle$ and $|\tilde{\boldsymbol{\Psi}}_0\rangle$ (top and bottom row, respectively) and different non-local interactions $g_J=\{0,3,5\}$ (appropriate columns from left to right). Note that strong enhancement of the non-spreading behavior takes place for stronger $g_J$ and  for a smeared out   initial state.}
 \label{fig2}
\end{figure}

We perform a  semi-classical evolution of the system governed by  Eqns.~\eqref{DavidovEq}, which allows us to observe spreading or non-spreading evolution of an effective width of the spatial density profile as a function of the parameters $\{ \omega_0,g_W,g_J \}$. The results can be visualized by the phase diagrams presented in Figs.~\ref{fig1} and~\ref{fig2}. These diagrams are obtained by plotting the maximal value of the $\sigma(t)$ reached during the evolution up to maximal time $T_{\mathrm{max}}=10$.  In order to avoid interference effects during the evolution caused by the boundaries,  all calculations are performed with a sufficiently large lattice of   $N=50$ lattice sites and with periodic boundary conditions. 
 We checked that the numerical results are insensitive to enlarging $N$ on the time-scales of study, and therefore the obtained results are also valid for infinitely large systems. Moreover, the chosen lattice size is similar to current experimental efforts in this direction.
First,  we focus on the case of a completely localized initial state $|\boldsymbol{\Psi}_0\rangle$ for $g_J=0$ (Fig.~\ref{fig1}).  
We qualitatively indicate five different regions on   the  phase diagram (left panel): {\it (I)} and {\it (II)} where the excitation is dressed by a cloud of vibrations and the excitation does spread; {\it (III)} where due to an exponential reduction of the hopping amplitude the excitation is localized in its initial position \cite{Fesser1982}; {\it (IV)} where  the sum of  vibration energy and exciton-vibration coupling is  larger than the hopping  energy,  giving rise to Davydov-like soliton behavior; {\it (V)} where $g_W\gtrsim\omega_0$ corresponding  to the  Discrete Breathers-like behavior  \cite{Flach1998,Juanico2007}. Distinct behavior of the system is also visible in these selected areas in the time evolution of $\sigma(t)$  (right panel of Fig.~\ref{fig1}).
This picture can be generalized to non-vanishing coupling $g_J$, which we investigate for the the second initial state $|\tilde{\boldsymbol{\Psi}}_0\rangle$ (Fig.~\ref{fig2}).
As can be seen, a slight delocalization of the initial state together with non-local coupling $g_J$ dramatically enhance the non-spreading behavior of the wave packet. It is a direct consequence of the non-local terms in \eqref{DavidovEq}.
\subsection*{Numerically exact approach}
\begin{figure}
\includegraphics[scale=2]{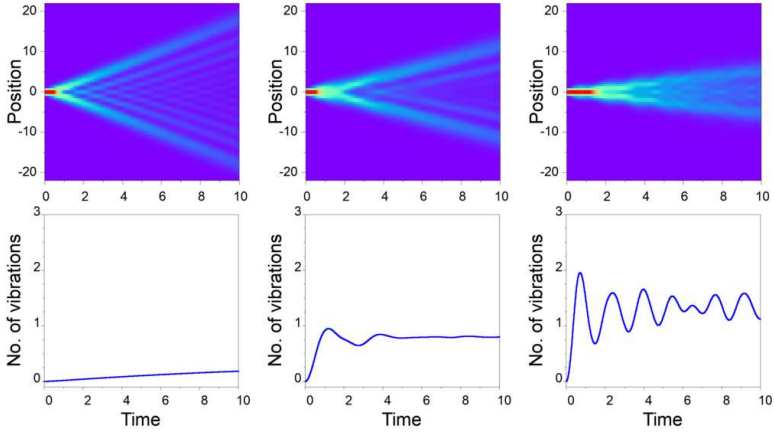} 
\caption{Exact evolution of the effective width of the spatial density profile governed by the Hamiltonian \eqref{Hamiltonian_final} for the initial state $|\boldsymbol{\Psi}_0\rangle$.  The bottom row shows the total number of vibrations $\hat{\cal N}_\mathtt{vib}$ created in the system during the evolution.  Consecutive columns correspond to different local couplings $g_W=\{0.1,0.75,1.5\}$. All calculations performed for $g_J=0$ and $\omega_0=3$. Note that for stronger interactions evident stabilization of the excitation density profile, along with increasing number of created vibrations, is observed. \label{fig3}}
\end{figure}

The results obtained  in the framework of the semi-classical  Davydov  approach can be supported by   numerically exact dynamics governed by the many-body Hamiltonian \eqref{Hamiltonian_final}. In this approach we represent Hamiltonian $\hat{\cal H}$ as a matrix in the Fock basis spanned by many-body states
% \begin{equation} \label{FockBasis}
% |i\rangle|m_1,\ldots,m_N\rangle = \hat{a}^\dagger_i(\hat{b}^\dagger_1)^{m_1}\cdots(\hat{b}^\dagger_N)^{m_N} |\mathtt{vac}\rangle,
% \end{equation}
$|i\rangle|m_1,\ldots,m_N\rangle = \hat{a}^\dagger_i(\hat{b}^\dagger_1)^{m_1}\cdots(\hat{b}^\dagger_N)^{m_N} |\mathtt{vac}\rangle$,
 {\it i.e.}, states with an excitation located exactly at site $i$ and with selected vibrational states $m_i$ for all sites. An arbitrary state of the system can be expressed as an appropriate superposition of the basis states. Since the operator of a total number of vibrations in the system $
\hat{\cal N}_\mathrm{vib}=\sum_i \hat{b}_i^\dagger\hat{b}_i
$
does not commute with the Hamiltonian \eqref{Hamiltonian_final},  therefore  an exact evolution is obtained only in the limit  where  all Fock states are taken into account. In practice, for numerical purposes, we assume that the total number of excitations cannot be larger than some well defined cut-off $M$. Then the results are treated as exact if increasing $M$ does not change the outcome noticeably \cite{Sowinski2,Dobrzyniecki}. Therefore,  for a given $M$, one can perform calculations only for a  small range of parameters for which creations of vibrations is limited. It is worth noticing that numerical complexity grows exponentially with the cut-off $M$. For our parameters, $N=50$ and $M=3$ the size of the corresponding Hilbert space exceeds 1.1 million. Other approaches based on quazi-exact dynamics are presented in \cite{Trugman2013,Trugman2015}.
As already expected from Fig.(\ref{fig2}), the numerically exact approach confirms that the on-site coupling ($g_W\ne 0)$ plays a dominant role in the excitation stabilization process. Therefore, without loss of generality, we consider now the minimal-coupling scenario with $g_J=0$
In Fig.~\ref{fig3} we show the time evolution of an initially localized excitation for $\omega_0=3$ and for three different values of the local coupling parameter $g_W=\{0.1, 0.75, 1.5\}$. It is clearly visible that for larger $g_W$ the wave packet of the excitation becomes more stable and spreads less. This effect is directly reflected in the number of vibrational modes created, which can be seen in the bottom row of Fig.~\ref{fig3}.
One can observe that increasing fluctuations of the total vibrations in the system stabilize excitation.
Since we reached the limits of our computational method with this size of the Hilbert space, we cannot increase the coupling parameter further. From Fig.~\ref{fig3} it can be seen that the total number of vibrations for $g_W=2$ is close to the limiting cut-off. At the same time, however, this is a strong argument for employing a quantum simulator, such as proposed in this letter, to validate the predictions of the semi-classical approach.

\section{Experimental parameters} The numerical predictions for the model described by the Hamiltonian \eqref{Hamiltonian_final} are quite general. For a quantum simulator we consider $^{87}$Rb atoms confined in an optical lattice determined by lattice spacing $R_0=1\,\mathrm{\mu m}$ and $V_0=100 E_R$ (recoil energy $E_R=2\pi^2\hbar^2/mR_0^2$) \cite{Pohl2014}, \textit{i.e.}, the local trap frequency is equal to $6.2\,\mathrm{kHz}$. We assume Rydberg states with principal quantum number $n=50$  for which $C_3^{sp} = 3.224\,\mathrm{GHz\,\mu m^3}$ \cite{Buchler2016}. We choose the dressing parameters as $\alpha = 0.015$ and  $\varDelta/2=\varDelta_s = \varDelta_p = 2.5\,\mathrm{GHz}$. 
With these values, the system mimics the Hamiltonian \eqref{Hamiltonian_final} with dimensionless parameters $\omega_0 = 4.7$, $g_W = 5.6$, and $g_J = 5.6$. These parameters can be easily tuned since they strongly depend on the lattice spacing $R_0$ and on the set of laser detunings. In this way, a  large and interesting area of the phase diagrams presented in Fig.~\ref{fig2} can be covered.
The estimated lifetime of Rydberg atoms excited to states with $n=50$ is $\tau_S = 65\,\mathrm{\mu s}$ and $\tau_P = 86\,\mathrm{\mu s}$ \cite{Beterov2009}.  The effective lifetime of a Rydberg dressed state is scaled by a factor $\alpha^{-2}$ and is sufficiently long to observe the non-spreading excitation behavior.  It is worth noting that also other experimental realizations, based for example on Rydberg microtrap arrays \cite{Leung2011}, can be considered as proper candidates for simulating this system.

\section*{Summary} We show that a system of ultracold  Rydberg atoms confined in a one-dimensional optical lattice may serve as a dedicated quantum simulator for excitation-vibration dynamics, which is a subclass of polaron dynamics. Since effective parameters of the resulting model can be easily tuned, the system can be used to mimic transport of excitation in biologically active proteins and to perform full quantum mechanical tests of the semi-classical predictions.  
 The proposed scheme may serve as a platform to investigate the HSSH bi- and many-polaron system \cite{Berciu2017a,Berciu2017b}. In particular, the character of the bi-polaron interactions can be tuned from repulsive to attractive by the experimental control parameters $g_W$ and $g_J$. Finally, we note that also disorder effects in the HSSH Hamiltonian \cite{Berciu2009} can be studied by introducing incommensurate optical lattices.

%\bibliography{_bibtex}

 \bibliographystyle{abbrv}
 % \bibliographystyle{unsrtnat}

% \end{thebibliography}

\section*{Acknowledgements}
This work was supported by the (Polish) National Science Center Grant No. 2016/22/E/ST2/00555,   by the Foundation for Fundamental Research on Matter (FOM), by the Netherlands Organisation for Scientific Research (NWO), and by the European Union H2020 FET Proactive project RySQ (grant N. 640378). 

\section*{Author contributions statement}
M.P was a founder of the idea for this project, prepared analytical calculations and prepared most of the numerical simulations. T.S. prepared exact-evolution part of the numerical simulations.
M.P., T.S., and S.K. equally contributed to discussions and wrote the manuscript.

\section*{Additional information}

\textbf{Competing Interests} The authors declare that they have no competing interests.

\end{document}